\documentclass[aps,prl, twocolumn]{revtex4}

\usepackage{graphicx}
\usepackage{amssymb}
\usepackage{amsmath}
\usepackage{colordvi}
\usepackage{color}

\newcommand{\cE}{{\cal E}}

\newcommand{\rev}[1]{\textcolor{black}{#1}}

\begin{document}

\title{Multipole  quantum droplets  in quasi-one-dimensional asymmetric mixtures}

\author{Yaroslav V. Kartashov$^1$ and Dmitry A. Zezyulin$^2$}

\affiliation{$^1$Institute of Spectroscopy, Russian Academy of Sciences, Troitsk, Moscow, 108840, Russia\\\medskip
$^2$School of Physics and Engineering, ITMO University, St. Petersburg 197101, Russia}


\date{\today}

\begin{abstract}

We study quantum droplets emerging in a quasi-one-dimensional asymmetric mixture of two atomic species with different intra-component coupling constants. We find that such mixtures support a rich variety of multipole quantum droplets, where the macroscopic wavefunction of one component changes its sign and features distinctive multipole structure, while the wavefunction of another component does not have zeros. Such multipole droplets have no counterparts in the reduced single-component model  frequently used to describe symmetric one-dimensional mixtures. We study transformations of multipole states upon variation of the chemical potential of each component and demonstrate that quantum droplets can split into separated  fundamental states, transform into flat-top multipoles, or into multipole component coupled to flat-top state with several humps on it, akin to anti-dark solitons. Multipole quantum droplets described here are stable in large part of their existence domain. Our findings essentially broaden the family of quantum droplet states emerging in the beyond-meanfield regime and open the way for observation of such heterostructured states in Bose-Bose mixtures.

\end{abstract}

\maketitle

\paragraph{Introduction.}

Quantum droplets (QDs) in Bose-Bose mixtures emerge due to the balance between meanfield interactions and the Lee-Huang-Yang (LHY) correction \cite{LHY} to the meanfield energy \cite{Petrov2015}. An important peculiarity of QDs stems from different roles played by the LHY correction in settings of different dimensionality. For example, in three-dimensional (3D) case LHY correction provides the repulsive effect and stabilizes the atomic mixture against the meanfield collapse, while in effectively one-dimensional (1D) mixtures  \cite{Petrov2016, Parisi1, Parisi2} the  LHY correction corresponds to effective attraction, compensating the intra-species repulsion and enabling the formation of self-bound states even in free space \cite{Petrov2016}. Competition between meanfield and LHY nonlinearities leads to very unusual shape transformations and stability properties of QDs. Experimentally, QDs have been observed in dipolar Bose-Einstein condensates \cite{Schmitt2016, FerrierBarbut2016, Wachtler2016, Baillie2016, Chomaz2016, Baillie2018}, as well as in Bose-Bose mixtures consisting of atoms in different hyperfine states, which are characterized by unequal coupling constants \cite{Cheiney2018, Cabrera2018, Semeghini2018, collisions}, and in heteronuclear Bose-Bose mixtures characterized by different atomic masses in two species \cite{hetero, Cavicchioli2022}. Current progress in experiments with QDs and their theoretical treatment is described in recent reviews \cite{Luo2021, Bottcher2021}.

Theoretical description of Bose-Bose mixtures frequently assumes a symmetric mixture of two species with equal atomic masses and intra-component coupling constants allowing to derive a reduced single-component model, where both species are described by the same Gross-Pitaevskii-like equation \cite{Petrov2016, AstaMalo, Katsimiga2023, Katsimiga2023CM, Edmonds2023, Khan2022, Abdullaev, Otajonov}.  At the same time,  this assumption  drastically limits the set of available nonlinear states. On this reason, the exploration of new types of QDs in essentially \textit{two-component} and \textit{asymmetric} mixtures becomes particularly important, as this situation is most frequently encountered in experiments. Besides simple transformation of shapes of QD components encountered in asymmetric mixtures \cite{Mithun2020}, it was found that asymmetry can lead to instabilities and new types of states \cite{Singh2020, Kartashov2022, Otajonov2022, Gangwar2024} that do not appear in the scalar case. The variety of stable QDs is particularly rich in multidimensional settings, where LHY correction can stabilize not only fundamental \cite{Cabrera2018, Semeghini2018, Cheiney2018, hetero, collisions, Cavicchioli2022}, but also excited states, such as vortical and rotating QDs \cite{Li2018, Kartashov2018, Kartashov2020, Kartashov2019, Tengstrand2019, Dong2021, Dong2022}. External trapping potentials \cite{Liu2019, Katsimiga2023, Zezyulin2023} and periodic lattices \cite{Morera2020, Zhou2019, Dong2020, Kartashov2024, Zhang2019, Zheng2021, Pathak2022, Nie2023, Liu2022, Liu2023, Huang2023} further enrich the structure and evolution regimes of QDs.  

Surprisingly, but as it comes to 1D QDs, only the simplest fundamental (i.e., nodeless) asymmetric states have been reported so far in settings without lattices or other types of confining potentials. In this Letter, we show that the family of 1D QDs in asymmetric mixtures in free space is in fact  much richer  and   includes stable multipole states  which do not have counterparts in the one-component system. In particular, we present the families of dipole and tripole QDs, identify the ranges of their existence and stability on the plane of chemical potentials of both species, and demonstrate that such QDs exhibit rich shape transformations within their existence domains, ranging from transformation of multipole QDs into several separated fundamental states, to flat-top multipoles, or multipoles coupled with humps on a localized flat-top plateau, resembling anti-dark states.

\paragraph{Model.} We study an asymmetric quasi-1D Bose-Bose mixture, where both species have equal atomic masses, but different intra-species coupling constants. This situation corresponds, for example, to experiments with a mixture of $^{39}$K atoms in different hyperfine states  \cite{Cheiney2018, Cabrera2018, Semeghini2018,collisions}. Evolution of dimensionless wavefunctions $\psi_{1,2}(x,t)$ is governed by the system \cite{Petrov2016, Minardi, Kartashov2022}:
\begin{equation}
\label{eq:main}
i\frac{\partial \psi_{1,2}}{\partial t} = -\frac{1}{2} \frac{\partial^2 \psi_{1,2}}{\partial x^2} +  \frac{\partial E(n_1, n_2)}{\partial n_{1,2}} \psi_{1,2},
\end{equation}
where $n_{1,2} = |\psi_{1,2}|^2$ and the energy density  reads
\begin{eqnarray}
E(n_1, n_2) = \frac{(g_{1}^{1/2} n_1   - g_{2}^{1/2} n_2)^2}{2} - \frac{2}{3\pi} (g_{1} n_1 + g_{2} n_2)^{3/2} \nonumber\\[1mm]
+ \frac{\delta (g_{1} g_{2})^{1/2}}{(g_{1} + g_{2})^2} (g_{1}^{1/2} n_2  + g_{2}^{1/2}n_1)^2 .\qquad
\end{eqnarray}
Dimensionless coefficients $g_{1,2}>0$ characterize intra-species interactions in each component,   coefficient $g_{12}<0$ defines inter-species interactions, and   $\delta = g_{12} +  ( g_{1} g_{2})^{1/2}$. In   experiments  $\delta$ is positive and small. In our simulations we use \cite{Kartashov2022}: $g_{1} =  0.639$, $g_{2} = 2.269$ and $g_{12} = -1$ that corresponds to $\delta \approx 0.204$.

We search for asymmetric stationary states with generically different chemical potentials and spatial profiles $\psi_{1,2} = e^{-i\mu_{1,2} t} u_{1,2}(x)$, where the functions $u_{1,2}$ are real-valued and localized, i.e. $u_{1,2}\to 0$ as $x\to \pm \infty$. The system (\ref{eq:main}) is known to support fundamental (or monopole) droplets, where both functions $u_{1,2}(x)$ are nodeless \cite{Mithun2020}. In contrast, here we introduce \emph{multipole} droplets, where one of   functions $u_{1,2}(x)$ can have zeros. Specifically, we focus on dipole and tripole solutions, in which the first component $u_1$ has one or two  zeros, respectively, while the second component $u_2$ is nodeless. Such solutions can be obtained from Eq. (\ref{eq:main}) using Newton's method \rev{with a suitable initial guess which is necessary for numerical iterations to converge to a multipole state and not to a ground-state monopole one. We used   initial guesses   in the form of  a superposition of two or three (for  dipoles and tripoles, respectively) well-separated sech-shaped profiles which were taken with the same sign  in the second component and with alternating  signs in the first component.
}

\begin{figure}
	\begin{center}
		\includegraphics[width=0.999\columnwidth]{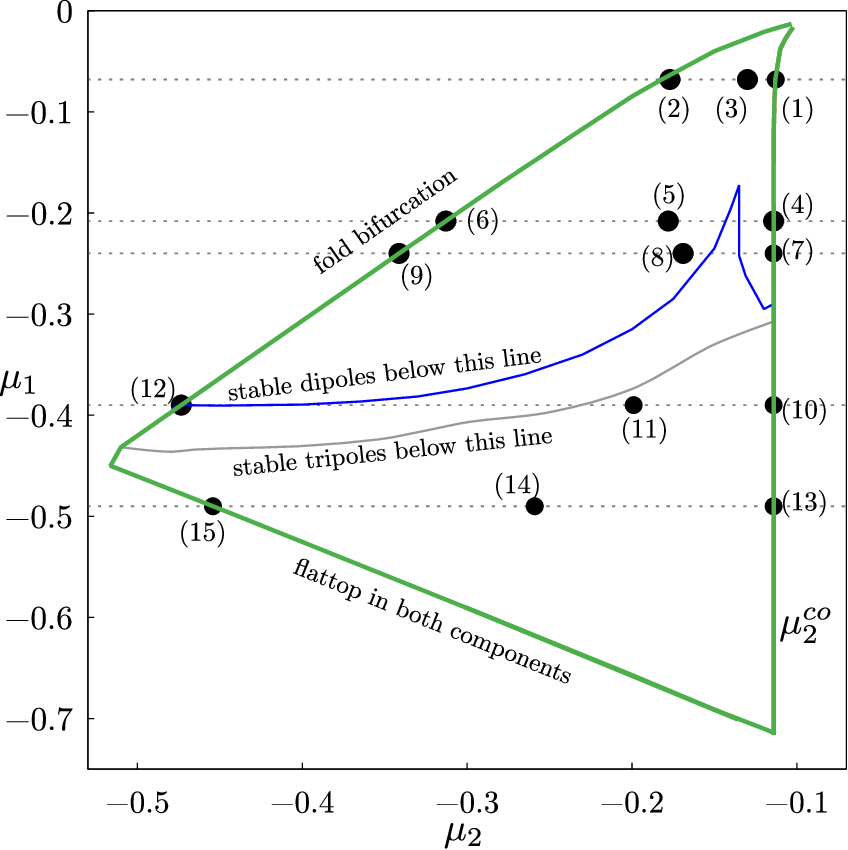}
	\end{center}
	\caption{The nearly-triangular area illustrates the  existence domain   for dipole and tripole QDs on the plane of chemical potentials $(\mu_1, \mu_2)$. The right boundary of the triangle corresponds to the cutoff value $\mu_2 = \mu_2^{co}$ of the single-component solution $(0, \psi_2)$. At the upper boundary of the triangle   dipole and tripole solutions undergo fold bifurcation, and at the lower boundary they transform into flat-top multipoles.  Enumerated circles correspond to specific solutions whose profiles are shown in Figs.~\ref{fig:profiles1} and \ref{fig:profiles2}. The diagram  displays two stability boundaries, with dipole and tripole QDs being stable below the corresponding boundary. \rev{In this  and subsequent figures all plotted quantities are   dimensionless.}} 
	\label{fig:existence}
\end{figure}   

\begin{figure}
	\begin{center}
		\includegraphics[width=1.00\columnwidth]{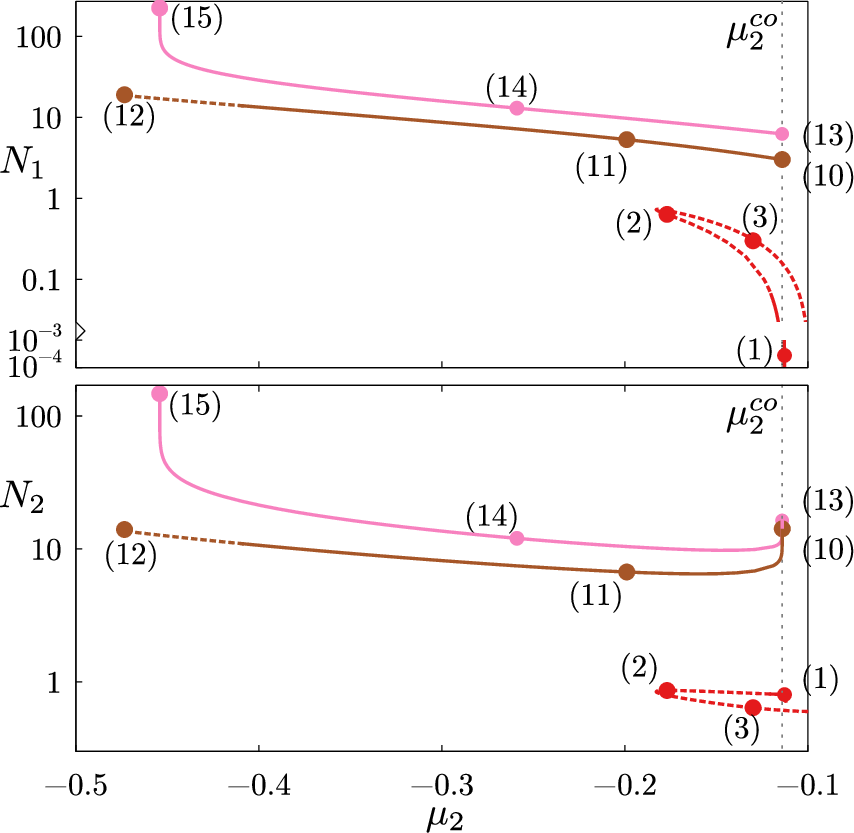}
	\end{center}
	\caption{Number of particles in the first (upper panel) and second (lower panel) component vs. $\mu_2$ for fixed $\mu_1$. Only dipole solutions are shown in this figure. Solid and dashed fragments correspond to stable and unstable solutions, respectively. Enumerated circles correspond to solutions marked in the existence diagram in Fig.~\ref{fig:existence} and plotted in Fig.~\ref{fig:profiles1}--\ref{fig:profiles2}. Red, brown, and pink lines correspond to  $\mu_1 = -0.068$; $\mu_1 = - 0.39$, and $\mu_1 =- 0.49$, respectively.}
	\label{fig:N1N2}
\end{figure} 

\begin{figure*}
	\begin{center}
 \includegraphics[width=1.0\textwidth]{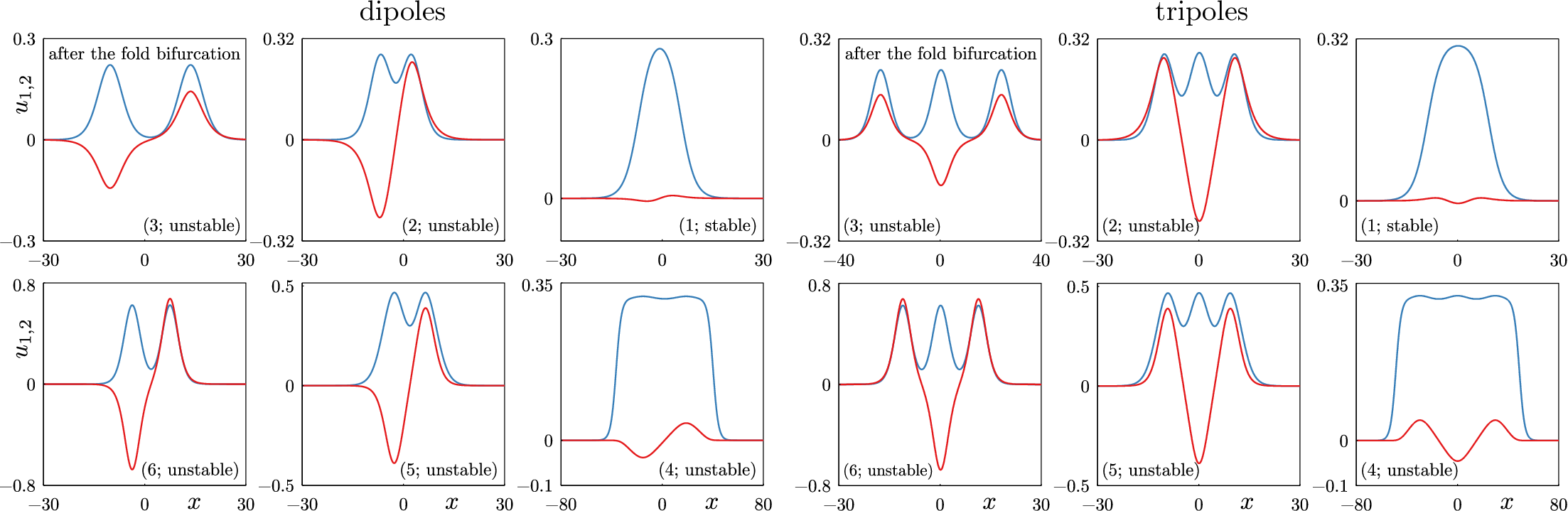}
	\end{center}
	\caption{Profiles $u_{1,2}(x)$ of two components of dipole and tripole QDs that transform into single-component state $(0, u_2)$ at the right boundary of the triangular existence area. Red and blue lines correspond to the first and second    components. Each row corresponds to the fixed value of $\mu_1$ (see horizontal dashed lines in Fig.~\ref{fig:existence}), while panels are enumerated in accordance with circles in the existence diagram in Fig.~\ref{fig:existence}. Notice that solution~3 is situated after the fold bifurcation, see corresponding solution in Fig.~\ref{fig:N1N2}.}
	\label{fig:profiles1}
\end{figure*}  

\begin{figure*}
	\begin{center}
		\includegraphics[width=1.0\textwidth]{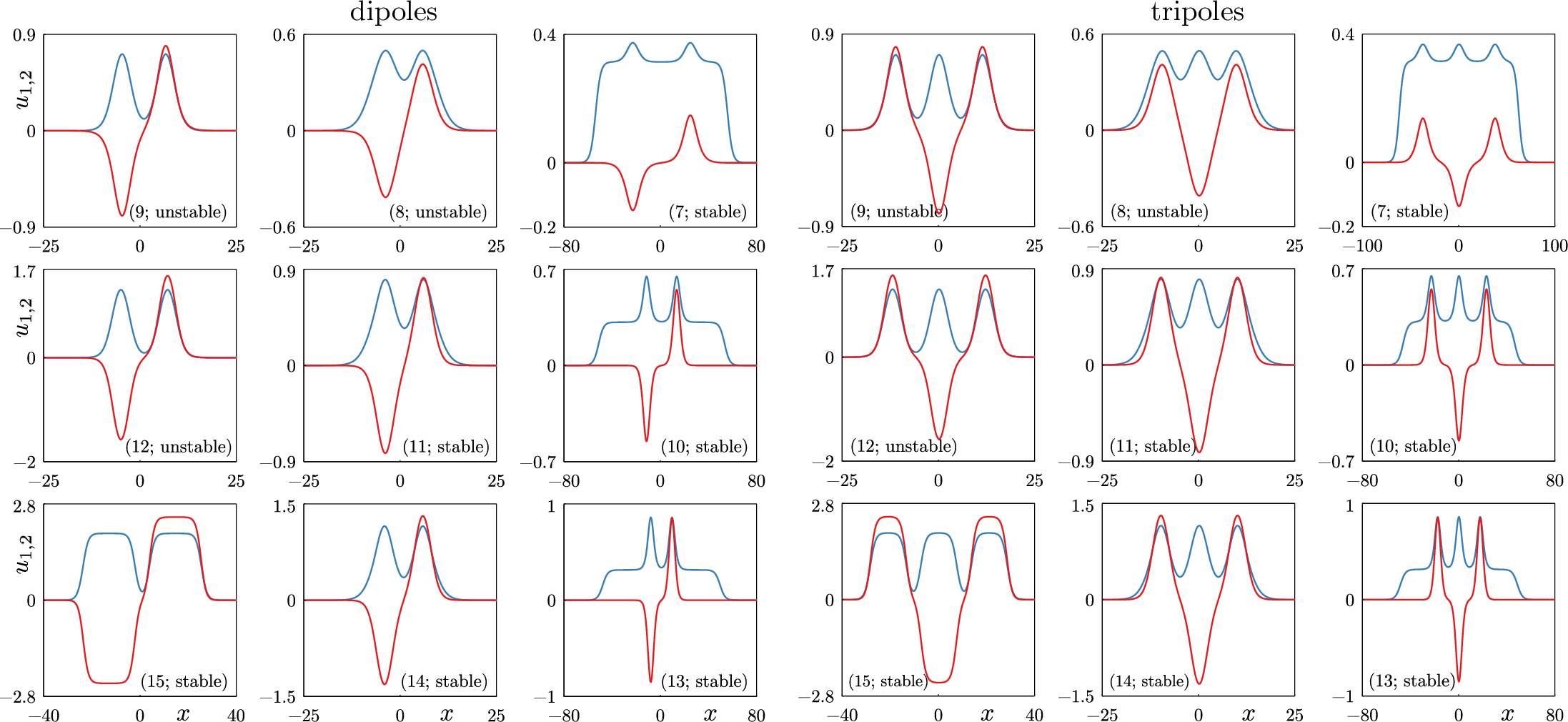}
	\end{center}
	\caption{Profiles $u_{1,2}(x)$ of dipole and tripole QDs that transform into multipole state in one component coupled to anti-dark state in other component at the right boundary of the triangular existence area. Red and blue lines correspond to the first and second components, respectively. Each row corresponds to the fixed value of $\mu_1$ and panels are enumerated in accordance with circles in the existence diagram in Fig.~\ref{fig:existence}.}
	\label{fig:profiles2}
\end{figure*}  

\paragraph{Classification of solutions.} First, we briefly describe the most important properties of a single-component model which can be formally obtained from (\ref{eq:main}) by setting the wavefunction of the first species to zero, i.e. $u_1\equiv 0$. Then the problem    reduces to single equation for $u_2(x)$ that has only monopole single-component solutions existing for chemical potential $\mu_2$ in the interval $\mu_2^{co} < \mu_2 <0 $, where the cutoff (`$co$') value is given by
\begin{equation} 
\mu_2^{co} := {\partial E(0, n_2^{co})} / {\partial n_2},
\end{equation}
and the limiting density $n_2^{co}$ can be determined from the condition
\begin{equation}
\label{eq:1}
\ {\partial E(0, n_2^{co})} / {\partial n_2}  =  E(0,  n_2^{co}) /  n_2^{co}.
\end{equation} 
It is well-known \cite{Petrov2016,AstaMalo} that as $\mu_2$ approaches $\mu_2^{co} $ from the right, the solution $u_2(x)$ develops the flattop shape, while $\textrm{max}(n_2)\to n_2^{co}$. Irrespective of the number of condensed atoms, the density $n_2(x)$ does not exceed  $n_2^{co}$. As a result, the droplet drastically broadens, and the number of atoms diverges as $\mu_2$ approaches the cutoff value from the right: $\lim_{\mu_2 \to \mu_2^{co}+0} N_2 =\infty$, where by definition $N_{1,2} = \int_{-\infty}^{\infty} |\psi_{1,2}|^2 dx$. When $\mu_2\to 0$, the norm $N_2$ vanishes. For the one-component system one gets
\begin{equation}
\label{eq:muco}
u_2^{co} = {2 g_{22}^{3/2}}/{(3\pi G_{2})} \quad \mbox{and} \quad\mu_2^{co} = - {2 g_{22}^3}/{(9\pi^2 G_{2})}, 
\end{equation}
where $G_{2} = g_{2} + \ {2\delta g_{1}^{3/2}g_{2}^{1/2}} / {(g_{1} + g_{2})^2}$.
For coupling constants adopted in our study, Eqs.~(\ref{eq:muco})  yield  $\mu_2^{co} \approx -0.114$ and $u_2^{co} \approx 0.315$. Remarkably, multipole two-component states described below exist   with chemical potentials situated to the left    of the  region of existence of single-component states.

Next, we proceed to the main part of our study devoted to    two-component mixtures with  $u_{1,2} \not \equiv 0$. In Fig.~\ref{fig:existence} we illustrate the domain of existence of dipole and tripole solutions on the plane $(\mu_1, \mu_2)$, while in Fig.~\ref{fig:N1N2} we plot representative dependencies of   numbers of particles $N_1$ and $N_2$ on the chemical potential of the second species $\mu_2$ for several fixed $\mu_1$ values. Representative profiles of multipole  QDs corresponding to the dots in Fig.~\ref{fig:existence} are displayed in Fig.~\ref{fig:profiles1} and \ref{fig:profiles2}. 

Multipole states exist within bounded domain on the $(\mu_1,\mu_2)$ plane that has nearly triangular shape, see Fig.~\ref{fig:existence}. The right side of the triangle approximately coincides with the $\mu_2 = \mu_2^{co}$ value corresponding to the cutoff for single-component condensate introduced above in Eq.~(\ref{eq:muco}).   We classify the found solutions in two groups which differ by their behavior near the right boundary of the existence area. As $\mu_2$ approaches $\mu_2^{co}$ from the left,   solutions of  the first group transform into single-component states $(0, u_2)$ briefly described above. Respectively, close to the right boundary of the existence area each solution of this type represents a localized nodeless state in the second component coupled to a small-amplitude multipole state in the first component, see examples   shown in Fig.~\ref{fig:profiles1} (specifically, solutions with numbers~1 and~4). Each solution of this type can be continued by decreasing $\mu_2$, while keeping $\mu_1$ fixed. This is accompanied by the increase of the amplitude of the first component which eventually becomes comparable with the second component, see solutions~2 and~5 in Fig.~\ref{fig:profiles1}. When $\mu_2$ reaches the left side of the triangular existence area, a fold bifurcation takes place, i.e., the solution family makes a U-turn  and continues towards \textit{increasing} values of $\mu_2$. A representative example of this fold bifurcation is presented in Fig.~\ref{fig:N1N2}, see the red curves corresponding to $\mu_1=-0.068$. After the fold bifurcation, with  the increase of $\mu_2$ such states gradually transform into well-separated sets of two (for dipoles) or three (for tripoles) fundamental QDs (see solutions with numbers~3 in Fig.~\ref{fig:profiles1} and the corresponding point in Fig.~\ref{fig:N1N2}).

The second group of solutions (illustrated in Fig.~\ref{fig:profiles2}) features   different behavior near the right boundary of the existence domain. Such states transform into a structure that comprises two (for dipoles) or three (for tripoles) out-of-phase well-separated humps in the first component coupled to two (or three) in-phase humps situated on the flattop pedestal in the second component, akin to anti-dark states. Examples of such  states are shown in Fig.~\ref{fig:profiles2} as solutions~7, 10, and 13.  As  $\mu_2$ approaches the right boundary of the existence area, the distance between the out-of-phase (in-phase) humps in the first (second) component increases. Therefore the number of particles diverges in the second component, but remains finite and nonzero in the first component. The difference between solutions from the first and second groups of solutions is best visible in the $N_1(\mu_2)$ dependencies plotted in upper panel of Fig.~\ref{fig:N1N2}. For solutions of the first group the number of particles $N_1$ vanishes at $\mu_2 \to \mu_2^{co}$ [see point~1 in Fig.~\ref{fig:N1N2}(a)], while for solutions of the second group $N_1$   remains nonzero [points~10 and 13 in Fig.~\ref{fig:N1N2}(a)]. As $\mu_2$ decreases, the behavior of QDs from the second group can be different depending on the value of chemical potential $\mu_1$ of the first component. If the branch of solutions reaches the left upper boundary of the triangle (see solutions 9 and 12 in Fig.~\ref{fig:existence} and \ref{fig:profiles2}), then  fold bifurcation takes place, by analogy with states from the first group.
In contrast, if $\mu_1$ is such that the family of states reaches the lower left border of the triangle, then both components develop flat-top shapes, while multipole structure is maintained (see dipole and tripole solutions~15  in Fig.~\ref{fig:existence} and  \ref{fig:profiles2}). We note that the boundaries of the existence domain on the $(\mu_1,\mu_2)$ plane for dipole and tripole QDs are identical.

\rev{Boundaries of the existence area in Fig.~\ref{fig:existence} show different response to the change of   intra-species coupling coefficient $g_{12}$ (and, respectively, to the change of   auxiliary coefficient $\delta$). Changing $g_{12}$ in the range from $-1.05$ to $-0.95$ (resp., $0.15\lesssim \delta \lesssim 0.25$), we found that the boundary corresponding to fold bifurcations does not change aprreciably, while the boundary corresponding to flattop shape in both components does change: for smaller values of $\delta$ the flattop regime  is achieved at smaller negative values of $\mu_1$ and $\mu_2$, i.e., the existence area broadens. \textit{Vice versa}, the increase of $\delta$ makes the existence domain narrower.}

\begin{figure}
	\begin{center}
		\includegraphics[width=0.999\columnwidth]{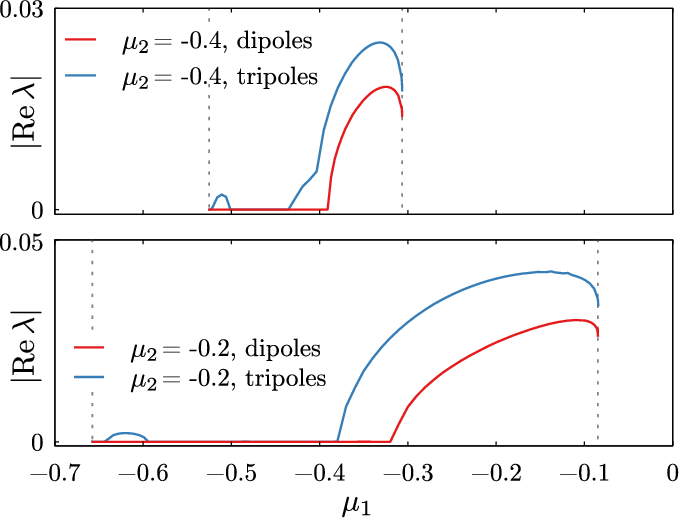}
	\end{center}
	\caption{Instability increments vs. chemical potential of the first component $\mu_1$ for dipole and tripole solutions at $\mu_2 = -0.2$ and $\mu_2=-0.4$. Horizontal spans of plotted curves are limited by the existence intervals of corresponding solutions (and additionally highlighted with vertical dashed lines). Apart from  relatively strong instability,   tripole QDs also feature narrow bands of weak instabilities, with increments below $ 10^{-3}$. These   weak instabilities are not shown in the existence diagram in Fig.~\ref{fig:existence}.}
	\label{fig:increments}
\end{figure}   

\begin{figure*}
	\begin{center}		
		\includegraphics[width=1.0\textwidth]{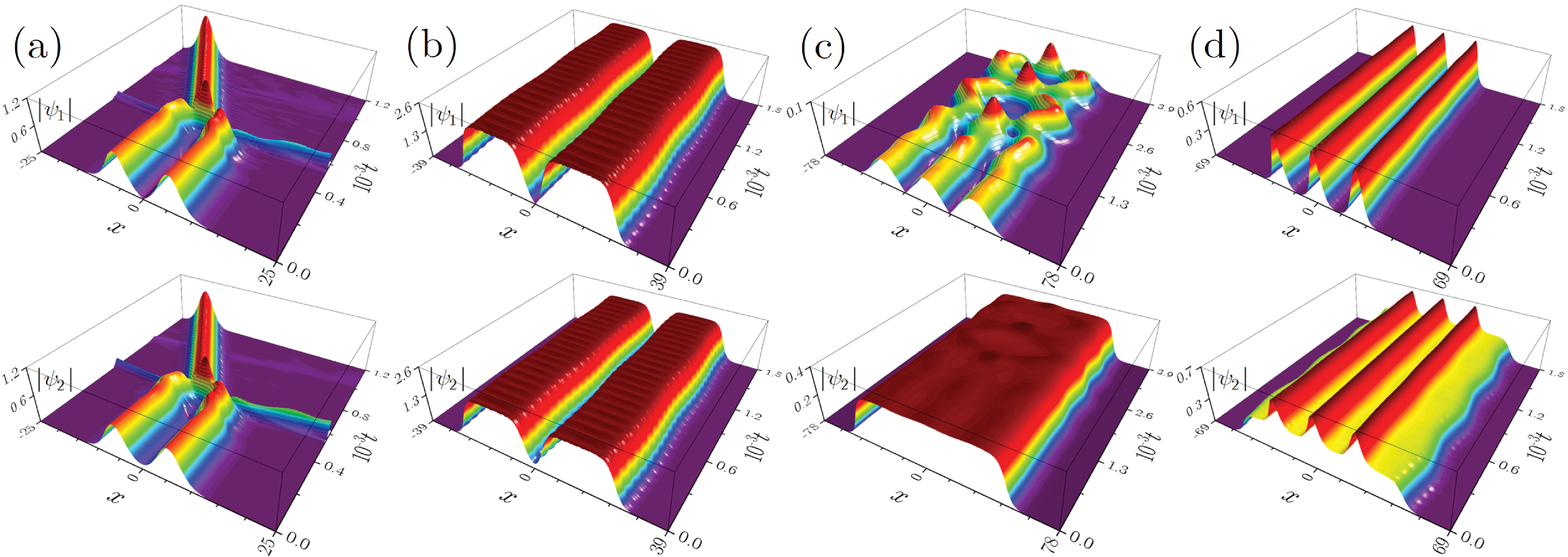}%
	\end{center}
	\caption{ (a,b) Dynamics of unstable and stable dipole QDs  which correspond to solutions with numbers 9 and 15 in Figs.~\ref{fig:existence}  and \ref{fig:profiles2}. (c,d) Dynamics of unstable and stable tripole QDs which correspond to solutions with numbers 4 and 10 in Figs.~\ref{fig:existence}, \ref{fig:profiles1}, and \ref{fig:profiles2}. First and second rows display amplitude of the first ($|\psi_1(x,t)|$) and second ($|\psi_2(x,t)|$) components, respectively.}
	\label{fig:dynamics}
\end{figure*}   

\paragraph{Stability.} To examine stability of   multipole droplets, we use    linear stability analysis and systematic dynamical simulations. For linear stability,  we take   perturbed   stationary solution  $\psi_{1,2} = e^{-i\mu_{1,2} t} [u_{1,2}(x) + \eta_{1,2}(x,t)]$, where $\eta_{1,2}$ are small perturbations. With representation   $\eta_{1,2}(x,t) = [P_{1,2}(x) + Q_{1,2}(x)] e^{\lambda t} + [ P_{1,2}^*(x) - Q_{1,2}^*(x)]e^{\lambda^*t}$,  the standard linearization procedure leads to      eigenvalue problem
\begin{eqnarray}
i\lambda P_{1,2} &=& \left[-\frac{1}{2} \frac{d^2\ }{dx^2}  - \mu_{1,2} + \frac{ \partial E}{\partial n_{1,2}}\right] Q_{1,2}, \\[3mm]%
i\lambda Q_{1,2} &=& \left[-\frac{1}{2} \frac{d^2\ }{dx^2}  - \mu_{1,2} + \frac{ \partial E }{\partial n_{1,2}} + 2u_{1,2}^2 \frac{\partial^2 E}{\partial n_{1,2}^2}\right] P_{1,2} \nonumber\\ &+& 2 u_1u_2 \frac{\partial^2 E}{\partial n_1 \partial n_2} P_{2,1}, 
\end{eqnarray}
where   partial derivatives of $E(n_1, n_2)$ are evaluated at $n_{1,2} =  u_{1,2}^2(x)$. This problem  has been solved numerically. For a given QD, the instability corresponds to perturbation with $\lambda$ having positive real part. Otherwise, i.e., if $\textrm{Re}(\lambda)\le 0$ for all perturbations, then the solution is stable. 

In spite of their complex internal structure, dipole and tripole states are stable in a wide subset of their existence domain. A simplified schematics illustrating  the main  boundaries between stable and unstable solutions  are presented in Fig.~\ref{fig:existence}: most of solutions of each type are stable below the corresponding line (blue for dipole QDs and gray for tripole QDs). In Fig.~\ref{fig:existence} we do not depict some narrow instability areas, where instability increment is rather small (of order $10^{-3}$ or below), because it is challenging to detect the borders of all such narrow domains. We have in addition found that the instability becomes strongly inhibited or even disappears completely for solutions situated close to the right boundary of the existence domain. Correspondingly, solutions~1 and 7 are stable, whereas solutions~4  feature only a ``mild'' dynamical instability, when time-dependent solutions  tend to maintain the internal structure at least for one of components, as opposed to strongly unstable states which completely lose their structure upon evolution. The difference between ``strong'' and ``mild'' instabilities is readily visible from examples of unstable dynamics presented in Fig.~\ref{fig:dynamics}: a strongly unstable dipole presented in Fig.~\ref{fig:dynamics}(a) dynamically transforms into a moving fundamental QD, while a weakly unstable tripole illustrated in Fig.~\ref{fig:dynamics}(c) preserves the shape of the second flattop component and displays quasiperiodic oscillations in the first component. Regarding  stable solutions,  illustrated in Fig.~\ref{fig:dynamics}(b,d),    initially introduced random perturbations lead  only to small-amplitude oscillations around   stationary shapes. 

The stability domain for tripoles is narrower than   for dipoles, as shown in Fig.~\ref{fig:existence}. In addition,  for tripoles there exists a  secondary instability band of finite width situated along the lower boundary of the triangular existence domain.  Instability increments in this band are approximately one order of magnitude smaller than those in the strong instability domain. For simplicity of the presentation, we   do not show   this additional instability area  in Fig.~\ref{fig:existence}; however, these instabilities are visible in Fig.~\ref{fig:increments}, where   instability increments are plotted  as   functions of chemical potential $\mu_1$ for two fixed values of $\mu_2$. At the lower-left boundary of the existence area (which corresponds to the flattop regime),   tripole solutions become stable again.

\begin{figure}
	\begin{center}
		\includegraphics[width=0.999\columnwidth]{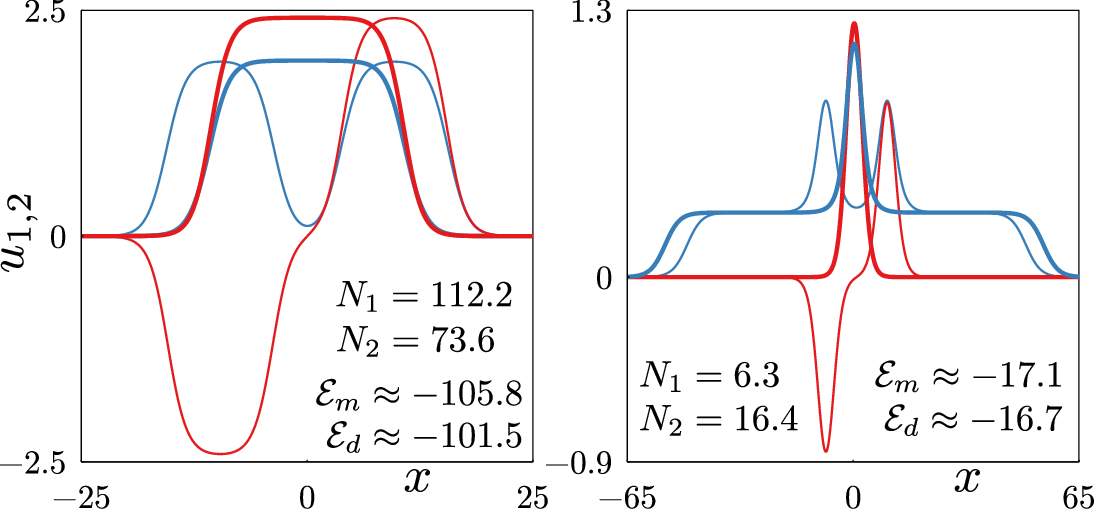}
	\end{center}
	\caption{\rev{Two examples of monopole (bold lines) and dipole (thin lines) QDs that share equal numbers of atoms in each species of the mixture. Numbers of atoms in the first ($N_1$) and second ($N_2$) species   and energies of monopole ($\cE_m$) and dipole ($\cE_d$) solutions are indicated in the plots. Red and blue lines correspond to profiles of the  first ($u_1$) and second  ($u_2$)   components.}  }
	\label{fig:mono}
\end{figure}

\rev{\paragraph{Comparison with monopole solutions.} For   completeness  we briefly discuss monopole, i.e., nodeless QDs that exist in the mixture with the adopted values of   coupling constants. In contrast to multipole QDs, monopole  solutions do not undergo   fold bifurcations indicated  in Fig.~\ref{fig:existence}.  Hence  for monopoles the corresponding boundary in absent in  Fig.~\ref{fig:existence}, and the  domain of their existence on the plane $(\mu_1, \mu_2)$  is broader. Computing    energy of steady states   $\cE = \int_{-\infty}^\infty [(\partial_x u_1)^2/2 + (\partial_x u_2)^2/2  + E(u_1^2, u_2^2)]dx$  for several representative solutions (two of which are presented in Fig.~\ref{fig:mono}), we   observe that for a mixture with fixed numbers of atoms in each species inequalities $\cE_m < \cE_d<\cE_t$ hold, where subscripts `$m$', `$d$', and `$t$' stay, respectively, for monopole, dipole, and tripole QDs. Hence in the energy space monopole  and multipole solutions represent ground and excited states, respectively.}

\paragraph{To conclude, } we  presented a previously unexplored class of multipole quantum droplets in quasi-1D asymmetric Bose-Bose mixtures. Such multipole solutions have no counterparts in the scalar model for symmetric system and   undergo unusual shape transformations within their existence domain. A particularly interesting class of   solutions consists of multiple out-of-phase humps in one component coupled to anti-dark state in another component. Despite their complex shapes,   higher-order excited states   are   stable in considerable part of their existence domain.

\begin{acknowledgments}
\paragraph{Acknowledgements.} Y.V.K. acknowledges funding by the research project FFUU-2024-0003 of the Institute of Spectroscopy of the Russian Academy of Sciences. The work of D.A.Z. was supported by the Priority 2030 Federal Academic Leadership Program.
\end{acknowledgments}

\end{document}